\documentclass[fleqn,usenatbib]{mnras}
\usepackage{newtxtext,newtxmath}
\usepackage[T1]{fontenc}
\usepackage{ae,aecompl}
\usepackage{graphicx}	
\usepackage{amsmath}	
\usepackage{amssymb}	
\usepackage{adjustbox}
\usepackage{float}
\usepackage{csquotes}
\usepackage{placeins}
\usepackage{adjustbox}

\usepackage{booktabs}

\title[Ionization profile of GRS 1915+105]{Exploring the radial disc ionization profile of the black hole X-ray binary GRS 1915+105}

\author[Shreeram \& Ingram]{
Soumya Shreeram$^1$
\& Adam Ingram$^1$\thanks{E-mail: adam.ingram@physics.ox.ac.uk}
\\
$^1$Department of Physics, University of Oxford, Keble Road, Oxford, OX1 3RH, UK.\\
}

\date{Accepted 2019 December 4. Received 2019 December 4; in original form 2019 April 25}

\pubyear{2018}

\begin{document}
\label{firstpage}
\pagerange{\pageref{firstpage}--\pageref{lastpage}}
\maketitle

\begin{abstract}

Accreting black holes show characteristic `reflection' features in their X-ray spectra, including the iron K$\alpha$ fluorescence line, which result from X-rays radiated by a compact central corona being reprocessed in the accretion disc atmosphere. The observed line profile is distorted by relativistic effects, providing a diagnostic for disc geometry. Nearly all previous X-ray reflection spectroscopy studies have made the simplifying assumption that the disc ionization state is independent of radius in order to calculate the restframe reflection spectrum. However, this is unlikely to be the case in reality, since the irradiating flux should drop off steeply with radius. Here we analyse a \textit{Nuclear Spectroscopic Telescope ARray} observation of GRS 1915+105 that exhibits strong reflection features. We find that using a self-consistently calculated radial ionization profile returns a better fit than assuming constant ionization. Our results are consistent with the inner disc being radiation pressure dominated, as is expected from the high inferred accretion rate for this observation. We also find that the assumed ionization profile impacts on the best fitting disc inner radius. This implies that the black hole spin values previously inferred for active galactic nuclei and X-ray binaries by equating the disc inner radius with the innermost stable circular orbit may be changed significantly by the inclusion of a self-consistent ionization profile.
\end{abstract}


\section{Introduction}

Stellar-mass black holes in X-ray binary systems (XRBs) and supermassive black holes (SMBHs) in active galactic nuclei (AGN) are thought to accrete material via a geometrically thin, optically thick accretion disc \citep{Shakura1973} that emits a thermal spectrum and a \textit{corona} located close to the hole that emits a power-law X-ray spectrum \citep{Thorne1975}. The exact geometry of the corona is still debated, with models including a magnetically supported layer above the disc \citep{Galeev1979}, a standing shock at the jet base \citep{Miyamoto1991,Fender1999} and a thick accretion flow formed by evaporation of the inner disc \citep[\textit{the truncated disc model};][]{Eardley1975,Done2007}.
X-rays from the corona irradiate and are reprocessed by the disc, imprinting characteristic `reflection' features onto the observed spectrum including a $\sim 6.4$ keV iron K$\alpha$ fluorescence line and a broad Compton scattering feature (`the Compton hump') peaking at $\sim 20-30$ keV \citep{Matt1991,Garcia2010}.
The observed shape of the reflection spectrum is distorted by relativistic orbital disc motion and gravitational redshift, leading to a relativistically broadened and skewed iron line profile \citep{Fabian1989}.
Fitting relativistic reflection models to the observed spectrum can therefore constrain the disc inner radius, $r_{\rm in}$, and consequently the dimensionless spin parameter of the black hole, $a=J/(M c r_g)$, where $J$ and $M$ are respectively the angular momentum and mass of the hole and $r_g=GM/c^2$. This is because a theoretical lower limit for the disc inner radius is provided by the innermost circular orbit (ISCO) of the black hole, which is a monotonic function of spin: $r_{\rm isco}=9~r_g$ for $a=-1$ (maximally retrograde) to $r_{\rm isco}=r_g$ for $a=1$ (maximally prograde).

This method has returned mainly estimates of rapid spin for AGN \citep{Reynolds2013}, perhaps favouring prolonged over chaotic accretion as the dominant mode of SMBH growth \citep{Fanidakis2011}.
In XRBs however, although there is broad consensus that $r_{\rm in}=r_{\rm isco}$ in the \textit{soft state} \citep[e.g.][]{Kubota2001,McClintock2014}, during which the spectrum is dominated by the thermal disc component, the iron line method has often indicated that $r_{\rm in} > r_{\rm isco}$ in the \textit{hard state}, during which the spectrum is dominated by the power-law component. Moreover, $r_{\rm in}$ is often measured to \textit{reduce} during the $\sim$weeks-months it takes to transition from the hard state to the soft state \citep[e.g.][]{Plant2015,Garcia2015}.

Over the years, increasingly sophisticated models for the rest frame reflection spectrum have been employed in reflection studies \citep{Lightman1980,George1991,Garcia2013}. An important parameter that governs the ionization of the disc atmosphere, and therefore the shape of the emergent reflection spectrum, is the ionization parameter
\begin{equation}
\xi(r)= 4\pi F_x(r)/n_e(r).
\label{eqn:xi}
\end{equation}
Here $F_x(r)$ is the $13.6$ eV to $13.6$ keV irradiating flux and $n_e(r)$ is the disc electron density. Almost all previous observational studies have assumed $\xi(r)=$constant. However, $F_x(r)$ is a steep function of radius, since the corona irradiates the inner disc more powerfully than the outer disc. It is thus very likely that ionization depends on disc radius. \citet{Svoboda2012} showed that considering a realistic radial ionization profile can explain the very steep emissivity profiles often required by spectral fits that parameterize the radial dependence of illuminating flux \citep[e.g.][]{Miller2013}, since the strength of the iron line increases with ionization and $\xi$ is expected to increase with proximity to the hole. Since the \textit{width} of the rest frame iron line also depends on ionization (to due e.g. Compton broadening and blending of lines for different ionization states), accounting for an ionization profile in reflection models may affect the resulting measurement of $r_{\rm in}$.

Here, we consider an observation of the XRB GRS 1915+105 made by the \textit{Nuclear Spectroscopic Telescope ARray} (\textit{NuSTAR}; \citealt{Harrison2013}) in 2012. We fit the spectrum with a reflection model that self-consistently calculates $F_x(r)$ assuming a point-like corona, and employ two different assumptions for $n_e(r)$ in order to calculate $\xi(r)$. We find that 1) the most physically motivated ionization profile that we consider, with $n_e(r)$ set by the \cite{Shakura1973} disc model, provides the best fit (preferred by $>4\sigma$ over constant ionization); and 2) the choice of ionization profile impacts the measured value of $r_{\rm in}$. Our best fit model returns $r_{\rm in}\approx 1.7~r_g$, contrasting with $r_{\rm in}\approx 6.7~r_g$ for the constant ionization model. This implies that the spin inferred by setting $r_{\rm in}=r_{\rm isco}$ is strongly affected by the assumed ionization profile. Section \ref{sec:DataReduction} details the \textit{NuSTAR} observation and our data reduction procedure. Section \ref{sec:models} describes all the models explored and the nomenclature used to describe them. We present our results in Section \ref{sec:results}, discuss them in Section \ref{sec:discussion} and conclude in Section \ref{sec:conclusions}.

\renewcommand{\tabcolsep}{0.8mm}
\begin{table*}

\centering
\begin{adjustbox}{width=\textwidth,center}
\begin{tabular}{|c|c|c|c|c|c|c|c|c|c|c|c|c|r|}
    \hline
    \textbf{\begin{tabular}[c]{@{}l@{}}$N_{\mathrm{H}}$\\ ($10^{22}$cm$^{-2}$)\end{tabular}} & \textbf{\begin{tabular}[c]{@{}l@{}}$T_{\mathrm{in}}$\\ (keV) \end{tabular}} &
    \textbf{q$_{\mathrm{in}}$} &
    \textbf{q$_{\mathrm{out}}$} &
    \textbf{\begin{tabular}[c]{@{}l@{}}$r_{\mathrm{br}}$\\ ($r_g$) \end{tabular}} &
    \textbf{\begin{tabular}[c]{@{}c@{}}a\\ \end{tabular}} &
    \textbf{\begin{tabular}[c]{@{}c@{}}$i$\\ (deg)\end{tabular}} &
    \textbf{$\Gamma$} &
    \textbf{$\log{\xi}$} &
    \textbf{$A_{\mathrm{Fe}}$} &
    \textbf{\begin{tabular}[c]{@{}c@{}}$E_{\mathrm{cut}}$\\ (keV)\end{tabular}} &
    \textbf{\begin{tabular}[c]{@{}l@{}}$f_{\mathrm{R}}$\\ \end{tabular}} &
    \textbf{\begin{tabular}[c]{@{}l@{}} $\log\xi_{\rm xill}$ \\ \end{tabular}} &
    \textbf{${\chi^2}/$d.o.f.} \\ \hline \hline
    $7.93_{-0.09}^{+0.23}$  & $0.40_{-0.002}^{+0.003}$   & $1.61_{-1.2}^{+2.4}$ & $>12$ & $1.65_{-0.02}^{+0.04}$ & $0.995_{-0.002}^{+0.001}$ & $77.29_{-0.1}^{+0.5}$  & $2.13_{-0.002}^{+0.03}$   & $3.06_{-0.02}^{+0.005}$  & $0.58_{-0.01}^{+0.01}$ & $79.4_{-0.36}^{+0.9}$ & $0.31_{-0.01}^{+0.02}$ & $2.29_{-0.1}^{+0.03}$ & 2708.74/2516  \\ \hline
\end{tabular}
\end{adjustbox}
\caption{Best fitting parameters from fitting the \textsc{relxill} model (Equation \ref{eqn:prevxspecmodel}). Here, $T_{\rm in}$ is the peak disc temperature, $i$ is the inclination angle (angle between the line of sight and the inner disc rotation axis), $A_{\rm Fe}$ is the disc iron abundance relative to solar, $f_{\rm R}$ is the reflection fraction (the system reflection fraction in the nomenclature of \citealt{Ingram2019}) and $\log\xi_{\rm xill}$ sets the ionization of the distant reflection. $\xi$ is in units of erg cm s$^{-1}$. All other parameters are described in the text. Following Z19, $r_{\rm in}$ was frozen to $r_{\rm isco}$. Uncertainties are $1\sigma$.
}
\label{tab:previous}
\end{table*}

\section{Data Reduction and Observation}
\label{sec:DataReduction}

\textit{NuSTAR} observed GRS 1915+105 on $3^{\rm rd}$ July 2012 when it was in a `plateau' ($\chi$) state. In this state, the X-ray flux and variability amplitude are relatively low, the spectrum is dominated by a hard power-law and none of the structured variability patterns seen in some other classes are present \citep{Belloni2000}. The $\chi$ state resembles the hard state seen from other X-ray binaries, in that the spectrum is hard, the variability amplitude is $\sim 20\%$, Type-C QPOs are present, and there is evidence for the presence of a compact radio jet \citep{Muno2001}.

We reduced the data using NuSTARDAS version 1.8.0 and the most up to date calibration files as of December 2017. We extracted cleaned event list files for focal plane modules (FPM) A and B with \texttt{nupipeline}. We then extracted FPMA and FPMB spectra from a $120''$ region centered on the source and generated response files using \texttt{nuproducts}. We extracted a background spectrum for each FPM from a $120''$ region far from the source and grouped the spectrum to have at least 25 counts in each grouped channel using the \texttt{ftool} \texttt{grppha}. Although the total elapsed time of the observation was $59.8$ ks, Earth occultations and various screenings bring the on source time down to $25.44$ ks and $25.56$ ks for the FPMA and FPMB respectively. Throughout, we use \textsc{xspec} v12.9 for spectral fitting.

\section{Models}
\label{sec:models}
\subsection{Previous analyses}

Previous analyses of this observation by \citet[][hereafter M13]{Miller2013} and \citet[][hereafter Z19]{zhang2019kerr} both used models that parameterize the radial emissivity as a twice broken power-law with inner index $q_{\rm in}$, outer index $q_{\rm out}$ and break radius $r_{\rm br}$. Both analyses assumed a constant ionization parameter and set $r_{\rm in}=r_{\rm isco}$ with $a$ left as a free parameter. The M13 spectral model consists of an exponentially cut-off power-law continuum (with photon index $\Gamma$ and cut off energy $E_{\rm cut}$) plus relativistic reflection. The restframe reflection spectrum is calculated using the model \textsc{reflionx} \citep{Ross2005}.

Z19 use the model \textsc{relxill\_nk} for their analysis. This model includes deviations from the Kerr metric through two deformation parameters, but reduces to the commonly used model \textsc{relxill} \citep{Dauser2013,Garcia2014} when the deformation parameters are set to zero and the Kerr metric is recovered. Since here we only consider the Kerr metric, the only Z19 fits that provide a basis for comparison are the ones with the deformation parameters frozen to zero. The restframe reflection spectrum is calculated using the model \textsc{xillver} \citep{Garcia2010,Garcia2013}, which includes more detailed atomic physics than \textsc{reflionx}. Z19 also additionally include a thermal disc component (\textsc{diskbb}) and a non-relativistic reflection component in their best fitting Kerr model (model 3' in their naming convention). This standalone \textsc{xillver} component likely represents enhanced distant reflection not accounted for in the relativistic model, perhaps from a flared outer disc, and its inclusion often significantly improves the fit (e.g. \citealt{Garcia2015,Ingram2016}). A subtlety accounted for in the \textsc{relxill} family of models, but not the model used by M13, is that the emergent restframe reflection spectrum depends on the initial trajectory of photons, which is in general different for photons that reach the observer from different parts of the disc due to light bending \citep{Garcia2014}.

Line-of-sight absorption is accounted for in both analyses with the multiplicative model \textsc{tbabs} \citep{Wilms2000}. The hydrogen column density, $N_H$, is the only parameter, and the abundances of all other elements relative to hydrogen are specified by a table. M13 use the abundances of \cite{Wilms2000}. Z19 on the other hand used the abundances of \cite{Anders1989} (Zhang, private communication). Here, we reproduce the result of Z19 by fitting the model
\begin{equation}
    \textsc{constant} \times \textsc{tbabs} \times \left( \textsc{diskbb} + \textsc{relxill} + \textsc{xillver}  \right),
    \label{eqn:prevxspecmodel}
\end{equation}
where the constant accounts for discrepancies in the absolute flux calibration between the FPMA and FPMB (frozen to unity for FPMA and the best fitting value is $\sim 1.02$ for FPMB). Following Z19, we include a direct disc component and a distant reflector with a free ionization parameter, and use the abundances of \cite{Anders1989}.

\subsection{Our Model}
 
In order to test the effect of including a realistic radial ionization profile on the best fitting parameter values and on the fit quality we consider the model
\begin{equation}
    \textsc{constant} \times \textsc{tbabs} \times \left( \textsc{diskbb} + \textsc{reltrans} + \textsc{xillver}  \right).
    \label{eqn:xspecmodel}
\end{equation}
The constant again accounts for calibration discrepancies, and we again allow the ionization parameter of the distant reflector to be free in the fit. For the relativistic reflection component, we use \textsc{reltrans} \citep{Ingram2019}. This model assumes an exponentially cut-off power-law illuminating spectrum, the restframe reflection spectrum is calculated using \textsc{xillver} and the angular dependence of the emergent reflection spectrum is self-consistently accounted for. The emissivity is not parameterized as a power-law function of radius. Instead it is calculated by ray tracing in the Kerr metric assuming that the illuminating corona is a stationary, point-like, isotropically radiating X-ray source located on the black hole spin axis a height $h$ above the hole, and that the disc is razor thin and in the black hole equatorial plane.
The relative normalization of the directly observed and reflected coronal emission is set by this calculation, except the reflection component is multiplied by a `boosting factor' $1/\mathcal{B}$.
This is a model parameter intended to approximately account for departures from the rather simple assumed coronal geometry.
For instance, $1/\mathcal{B}<1$ mimics the effect of the source moving away from the disc. \textsc{reltrans} can be set to consider a constant ionization parameter $\xi(r)=\xi_0$.
In this case, \textsc{reltrans} with $1/\mathcal{B}=1$ returns identical results to \textsc{relxilllp} with the setting \texttt{fixReflFrac}$=1$ \citep{Dauser2013,Ingram2019}.

\textsc{reltrans} also allows different ionization profiles to be employed (see Equation \ref{eqn:xi}). The illuminating flux $F_x(r)$ is known for the lamppost geometry, but the electron density $n_e(r)$ is uncertain. We test two assumptions. First, we assume that $n_e$=constant, giving $\xi(r) \propto F_x(r)$ (following \citealt{Svoboda2012,Kammoun2019}). We then consider the density profile in `zone A' of the \cite{Shakura1973} disc model, giving $\xi(r) \propto F_x(r)\ r^{-3/2}[1-\sqrt{r_{\rm in}/r}]^2$. This is the inner of three zones defined by \cite{Shakura1973}, where the pressure is radiation dominated and the opacity is dominated by electron scattering. For a bright source such as GRS 1915+105 the inner regions of the disc, which dominate the reflection emissivity, are likely in the zone A regime. In both cases, the peak ionization parameter is a model parameter and the calculation is discretized into 10 ionization zones (set by the environment variable {\fontfamily{cmtt}\selectfont ION\_ZONES}), which \cite{Ingram2019} found to be sufficient.

For the absorption model, we consider both the abundances of \cite{Anders1989} and those of \cite{Wilms2000}. Although the \cite{Wilms2000} abundances are the most up to date, we note that elemental abundances departing from those of the inter stellar medium have been observed for GRS 1915+105 in \textit{Chandra} grating spectra \citep{Lee2002}. A fraction of the foreground material may therefore be local to GRS 1915+105, or may have been enriched by outflows from the source over time. It is therefore not clear which set of abundances is neccesserily more appropriate, and so we trial both.

\begin{table}
\centering
    \begin{tabular}{|l|c|}
    \hline
    \textbf{Ionization profile} & \textbf{Label} \\ \hline \hline
    $\xi(r)=\xi_0$ & $A$ \\ \hline
    $\xi(r)\propto F_x(r)$ & $B$ \\ \hline    
    $\xi(r)\propto F_x(r)r^{-3/2}[1-\sqrt{r_{\rm in}/r}]^2$ & $C$ \\ \hline
    \end{tabular}
\caption{Naming convention for the ionization profiles used in this paper. Models are named using the above labels, and a subscript specifies whether the abundances of \citep{Wilms2000} or \citep{Anders1989} are assumed for the line of sight absorption model (subscript wilm or angr respectively). A prime indicates that the spin parameter is free in the fit.}
\label{Tab: models}
\end{table}
\begin{figure*}
 \centering
 \includegraphics[width=0.9\textwidth]{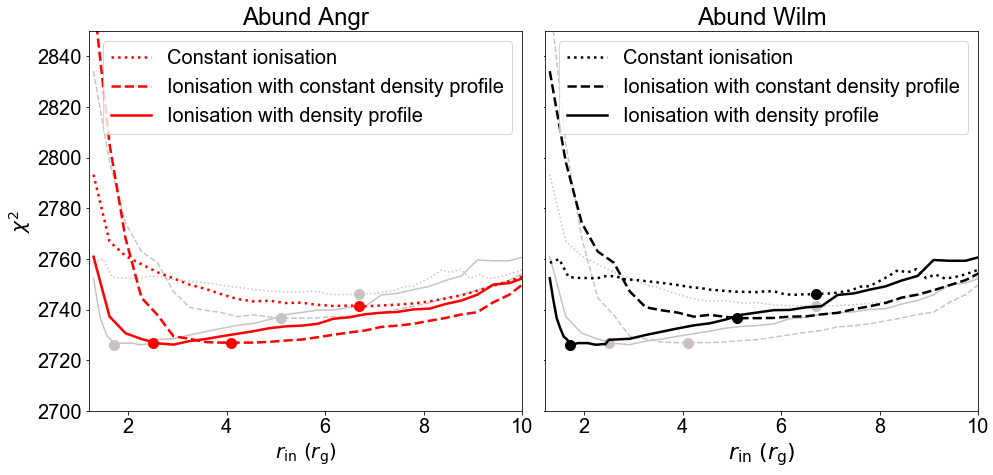}
 \vspace*{-1mm}
 \caption{$\chi^2$ versus $r_{\rm in}$ for \textsc{reltrans} models $A_{\rm angr}$ (red dotted), $B_{\rm angr}$ (red dashed), $C_{\rm angr}$ (red solid), $A_{\rm wilm}$ (black dotted), $B_{\rm wilm}$ (black dashed) and $C_{\rm wilm}$ (black solid). The filled circles mark the minimum $\chi^2$ for each model. We see that the assumed ionization profile changes both the minimum $\chi^2$ and the best fitting $r_{\rm in}$. The best fit is for the C models, which calculates the radial ionization profile by assuming a `zone A' Shakura \& Sunyaev radial density profile.}
 \label{fig: resultsPlot}
\end{figure*}

\begin{table*}
\centering
\resizebox{\textwidth}{!}{
\begin{tabular}{lcccccccccccccccr}
\hline
\textbf{Model} & \begin{tabular}[c]{@{}c@{}}$N_{\rm H}$\\ ($10^{-22} {\rm cm^{-2}}$)\end{tabular} & \begin{tabular}[c]{@{}c@{}}$T_{\rm in}$\\ (keV)\end{tabular} & \begin{tabular}[c]{@{}c@{}}$N_{\rm d}$\\ $(10^4)$\end{tabular} & \begin{tabular}[c]{@{}c@{}}$h$\\ ($r_{\rm g}$)\end{tabular} & \begin{tabular}[c]{@{}c@{}}$a$\\\end{tabular} & \begin{tabular}[c]{@{}c@{}}$i$\\ (deg)\end{tabular} & \textbf{\begin{tabular}[c]{@{}c@{}}$r_{\rm in}$\\ ($r_{\rm g}$)\end{tabular}} & $\Gamma$ & $\log\xi$ & \multicolumn{1}{l}{$\log\xi_{\rm xill}$} & $A_{\rm Fe}$ & \begin{tabular}[c]{@{}c@{}}$E_{\rm cut}$\\ (keV)\end{tabular} & $1/\mathcal{B}$ & $N_{\rm rel}$ & \begin{tabular}[c]{@{}c@{}}$N_{\rm xill}$\\ ($10^{-3}$)\end{tabular} & $\chi^2/$d.o.f. \\ \hline \hline
$A_{\rm angr}$ & $7.1_{-0.5}^{+0.5}$ & $0.42_{-0.02}^{+0.14}$ & $5.1_{-1.1}^{+1.5}$ & $1.90_{-0.10}^{+1.97}$ & $\equiv0.998$ & $61.7_{-9.7}^{+1.3}$ & $6.7_{-1.2}^{+1.1}$ & $1.93_{-0.05}^{+0.04}$ & $3.62_{-0.10}^{+0.09}$ & $2.76_{-0.05}^{+0.08}$ & $0.79_{-0.21}^{+0.49}$ & $52.8_{-4.5}^{+4.1}$ & $1.5_{-0.6}^{+1.5}$ & $0.83_{-0.38}^{+0.06}$ & $5.415_{-0.001}^{+0.003}$ & $2740.9/2518$ \\ \\ \hline
$A_{\rm wilm}$ & $11.7_{-0.8}^{+0.8}$ & $0.43_{-0.01}^{+0.01}$ & $5.0_{-0.9}^{+1.1}$ & $2.2_{\rm p}^{+2.5}$ & $\equiv0.998$ & $60.1_{-1.3}^{+1.4}$ & $6.7_{-1.3}^{+1.3}$ & $1.92_{-0.02}^{+0.04}$ & $3.76_{-0.04}^{+0.04}$ & $2.75_{-0.05}^{+0.07}$ & $0.84_{-0.12}^{+0.36}$ & $52.45_{-2.3}^{+1.6}$ & $1.5_{-0.4}^{\rm p}$ & $0.4_{-0.1}^{+0.1}$ & $4.6_{-1.8}^{+1.0}$ & $2744.0/2518$ \\ \\ \hline
$B_{\rm angr}$ & $7.5_{-0.5}^{+0.6}$ & $0.40_{-0.01}^{+0.02}$ & $7.5_{-1.7}^{+2.3}$ & $5.69_{-2.5}^{+2.8}$  &$\equiv0.998$ & $65.6_{-1.3}^{+1.3}$ & $4.1_{-1.7}^{+0.7}$ & $2.05_{-0.03}^{+0.03}$ & $4.3_{-0.9}^{+0.1}$ & $3.12_{-0.08}^{+0.13}$ & $0.5_{\rm p}^{+0.06}$ & $65.9_{-4.5}^{+2.8}$ & $1.1_{-0.4}^{+0.4}$ & $0.06_{-0.01}^{+0.04}$ & $9.6_{-1.5}^{+1.0}$ & $2727.2/2518$ \\ \\ \hline
$B_{\rm wilm}$ & $12.5_{-1.0}^{+0.8}$ & $0.42_{-0.01}^{+0.01}$ & $7.43_{-2.1}^{+0.9}$ & $4.1_{-1.1}^{+4.4}$ & $\equiv0.998$ & $64.2_{-1.6}^{+1.0}$ & $5.1_{-2.2}^{+0.7}$ & $2.07_{-0.03}^{+0.04}$ & $4.25_{-0.34}^{+0.08}$ & $3.1_{-0.3}^{+0.1}$ & $0.5_{\rm p}^{+0.03}$ & $71.4_{-2.1}^{+3.1}$ & $1.23_{-0.3}^{+0.1}$ & $0.08_{-0.02}^{+0.10}$ & $10.02_{-0.01}^{+0.01}$ & $2736.6/2518$ \\ \\ \hline
$C_{\rm angr}$ & $7.6_{-0.6}^{+0.5}$ & $0.40_{-0.01}^{+0.02}$ & $8.8_{-2.6}^{+2.6}$ & $7.7_{-3.5}^{+2.6}$ & $\equiv0.998$ & $66.4_{-1.7}^{+1.7}$ & $2.5_{-0.4}^{+0.8}$ & $2.06_{-0.05}^{+0.02}$ & $4.0_{-0.2}^{+0.3}$ & $3.2_{-0.1}^{+0.1}$ & $0.5_{\rm p}^{+0.08}$ & $63.7_{-4.8}^{+1.4}$ & $0.9_{-0.3}^{+0.2}$ & $0.050_{-0.005}^{+0.022}$ & $10.04_{-0.01}^{+0.01}$ & \textbf{$2725.2/2518$} \\ \\ \hline
$C_{\rm wilm}$ & $12.6_{-0.8}^{+1.0}$ & $0.40_{-0.01}^{+0.02}$ & $10.3_{-2.2}^{+3.0}$ & $12.5_{-3.5}^{+1.7}$ & $\equiv0.998$ & $67.1_{-2.3}^{+1.1}$ & $1.725_{-0.07}^{+0.64}$ & $2.11_{-0.04}^{+0.04}$ & $4.2_{-0.5}^{+0.1}$ & $3.21_{-0.12}^{+0.14}$ & $0.5_{\rm p}^{+0.04}$ & $66.6_{-2.1}^{+3.7}$ & $0.86_{-0.15}^{+0.13}$ & $0.052_{-0.006}^{+0.003}$ & $11.4_{-1.6}^{+1.5}$ & \textbf{$2725.9/2518$} \\  \\ \hline \hline
$C_{\rm wilm}^{'}$ & $12.7_{-0.6}^{+0.4}$ & $0.40_{-0.01}^{+0.01}$ & $10.0_{-0.1}^{+1.0}$ & $10.0_{-1.0}^{+1.9}$ & $0.976_{-0.021}^{p}$ & $67.1_{-1.1}^{+1.9}$ & $1.74_{-0.53}^{+0.32}$ & $2.11_{-0.02}^{+0.02}$ & $4.2_{-0.3}^{+0.3}$ & $3.16_{-0.14}^{+0.08}$ & $0.50_{-0.03}^{+0.03}$ & $66.4_{-1.7}^{+2.9}$ & $0.92_{-0.04}^{+0.03}$ & $0.050_{-0.001}^{+0.002}$ & $11.0_{-0.4}^{+0.6}$ & $2725.4/2517$ \\ \\ \hline \hline
\end{tabular} }
\caption{Best fitting model parameters. The model naming convention is described in Table \ref{Tab: models}. $T_{\rm in}$ and $N_{\rm d}$ are \textsc{diskbb} parameters (peak temperature and normalisation). $\log\xi$ is the maximum value that $\log\xi(r)$ reaches in the disc and $\log\xi_{\rm xill}$ is the ionization of the standalone \textsc{xillver} component. In both cases, $\xi$ is in units of erg cm s$^{-1}$. $1/\mathcal{B}$ is the boost parameter described in the text. $N_{\rm rel}$ and $N_{\rm xill}$ are the normalisation parameters of the \textsc{reltrans} and \textsc{xillver} components respectively. Symbols $\equiv$ and $p$ indicate respectively that a parameter is fixed or pegged to an extreme value.}
\label{tab:results2}
\end{table*}
\begin{table*}
\begin{tabular}{cccccccccccr}
\hline
\begin{tabular}[c]{@{}c@{}}$N_{\rm H}$\\ ($10^{-22} {\rm cm^{-2}}$)\end{tabular} & \begin{tabular}[c]{@{}c@{}}$h$\\ $r_{\rm g}$\end{tabular} & \begin{tabular}[c]{@{}c@{}}$i$\\ (deg)\end{tabular} & $a$ & \begin{tabular}[c]{@{}c@{}}$r_{\rm in}$\\ $r_{\rm g}$\end{tabular} & $\Gamma$ & $log{\xi}$ & $A_{\rm Fe}$ & \begin{tabular}[c]{@{}c@{}}$E_{\rm cut}$\\ $keV$\end{tabular} & $1/\mathcal{B}$ & \begin{tabular}[c]{@{}c@{}}$N$\\ ($10^{-2})$\end{tabular} & $\chi^2/\nu$ \\ \hline \hline
$4.9_{-0.1}^{+0.1}$ & $7.62_{-0.6}^{+4.3}$ & $17.9_{-3.2}^{+1.2}$ & $0.88_{-0.13}^{+0.06}$ & $22_{-8}^{+3}$ & $1.64_{-0.01}^{+0.01}$ & $4.3_{-0.1}^{+0.1}$ & $6.5_{-1.1}^{+1.0}$ & $36.4_{-0.2}^{+0.5}$ & $1.1_{-0.1}^{+0.1}$ & $3.64_{-0.1}^{+0.4}$ & $2838.3/2521$  \\ \hline
\end{tabular} 
\caption{Best fitting parameters of the \textsc{constant $\times$ reltrans} model without the \textsc{diskbb} and \textsc{xillver} components. The ionization profile $C_{\rm wilm}$, as described in Table \ref{Tab: models}, is implemented. All the parameters are described in the text.}
\label{tab:Z}
\end{table*}

\begin{figure}
    \includegraphics[width=\columnwidth,trim=1.3cm 1.8cm 3.0cm 1.8cm,clip=true]{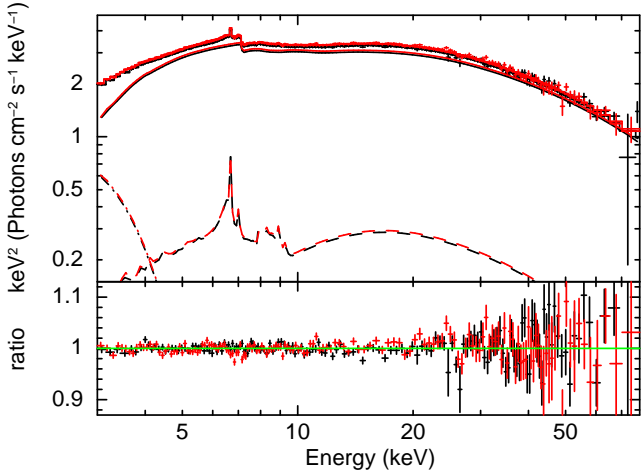}
    \caption{\textit{Top:} FMPA (black) and FPMB (red) spectrum unfolded around our best fitting model, $C_{\rm wilm}^{'}$ (ionization profile assuming Shakura and Sunyaev density profile). The components are \textsc{reltrans} (solid), \textsc{diskbb} (dot-dashed) and \textsc{xillver} (dashed). \textit{Bottom:} Ratio of data to model showing good agreement.
    Error bars are $1\sigma$ and $\chi^2/{\rm d.o.f.} = 2725.4/2518$.}
\label{fig: bestFitSpectrum}
\end{figure}
\begin{figure*}
\centering
\includegraphics[width=\textwidth,trim=1.0cm 2.0cm 1.0cm 2.0cm,clip=true]{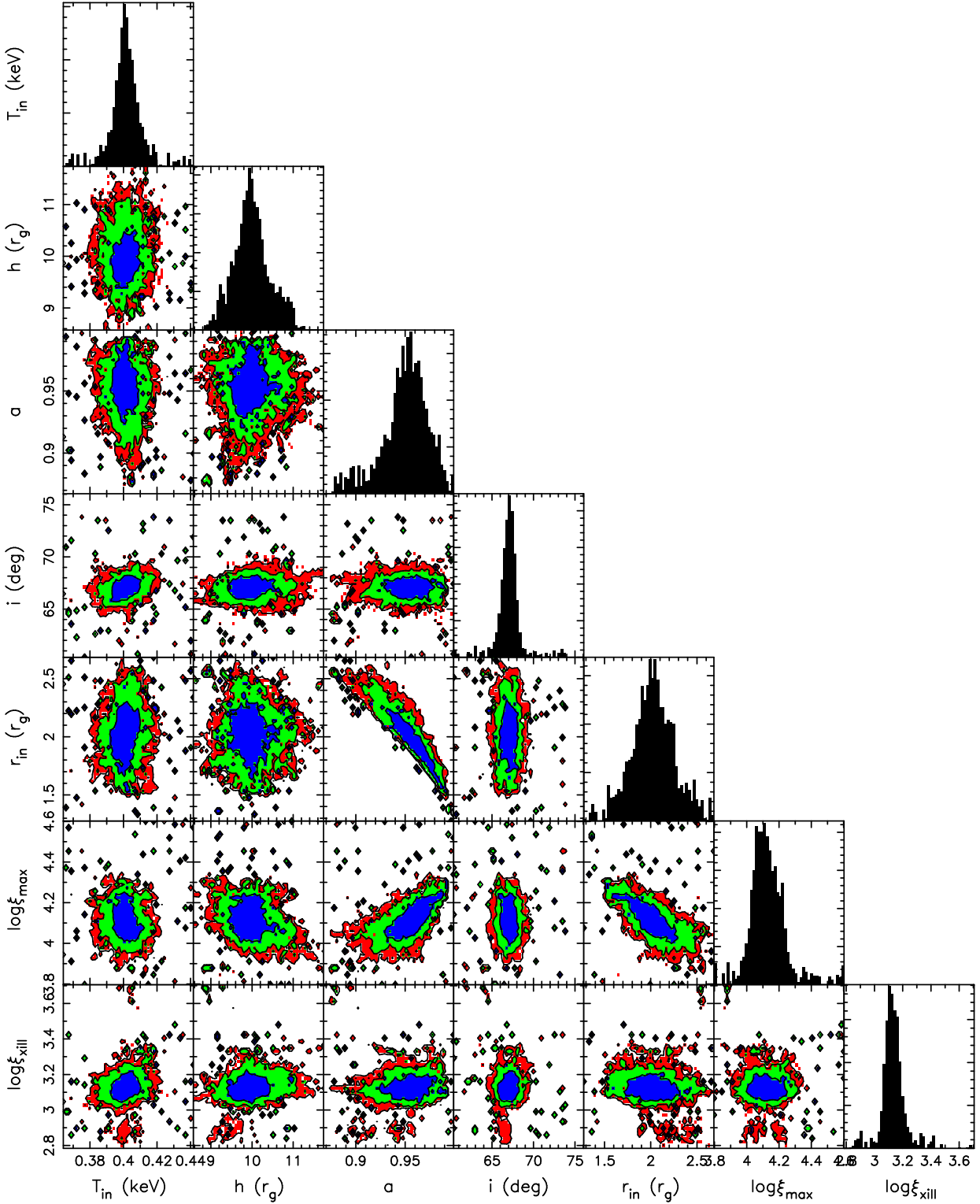}
\caption{Corner plot for our preferred model ($C_{\rm wilm}^{'}$). $1$, $2$ and $3~\sigma$ confidence contours are coloured blue, green and red respectively. The distributions are sampled using an MCMC simulation ran within \textsc{xspec} using the Goodman-Were algorithm with a total length of 307,200 steps spread over 256 walkers, after an initial burn-in period of 19,968 steps.}
\label{fig:corner}
\end{figure*}
\begin{figure}
\centering
\includegraphics[width=\columnwidth,trim=0.0cm 1.8cm 0.0cm 1.8cm,clip=true]{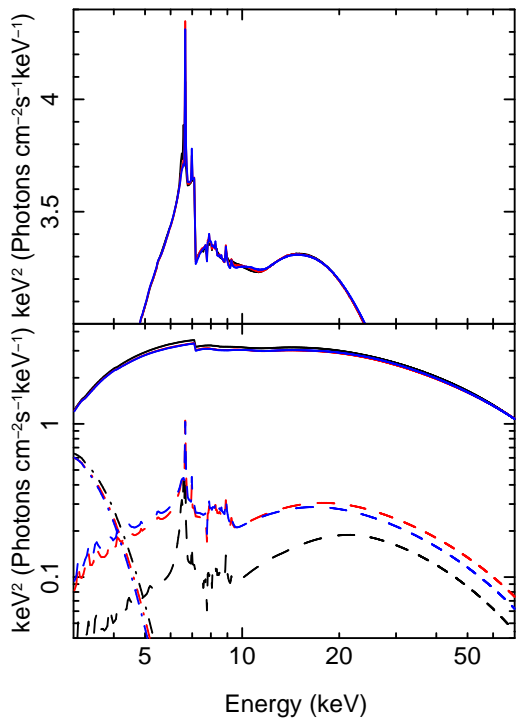}
\caption{Spectrum of models $A_{\rm wilm}$ (black), $B_{\rm wilm}$ (red) and $C_{\rm wilm}^{'}$ (blue). The top panel displays the full (FPMA) spectrum. The bottom panel displays the separate components: \textsc{reltrans} (solid), \textsc{diskbb} (dot-dashed) and \textsc{xillver} (dashed). The differences are subtle, but $C_{\rm wilm}^{'}$ is preferred over $A_{\rm wilm}$ with $>4\sigma$ confidence.}
\label{fig:eemo_compare}
\end{figure}

\begin{table}
\centering{
\begin{tabular}{|l|c|c|}
\hline
\multicolumn{1}{c}{\textbf{Model}} & \textbf{\begin{tabular}[c]{@{}c@{}}System \\ reflection fraction\end{tabular}} & \multicolumn{1}{c}{\textbf{\begin{tabular}[c]{@{}c@{}}Observer's \\ reflection fraction\end{tabular}}} \\ \hline \hline
$A_{\rm angr}$ & $0.57$ & $0.76$ \\
$A_{\rm wilm}$ & $0.59$ & $0.85$ \\
$B_{\rm angr}$ & $1.40$ & $1.32$ \\
$B_{\rm wilm}$ & $1.18$ & $1.30$ \\
$C_{\rm angr}$ & $1.51$ & $1.15$ \\
$C_{\rm wilm}$ & $1.39$ & $1.04$ \\
$C_{\rm wilm}'$ & $1.41$ & $1.07$ \\ \hline
\end{tabular}%
}
\caption{Summary of the `system reflection' fraction and `observer's reflection fraction' for the best fitting parameters of the seven models presented in Table \ref{tab:results2}.}
\label{tab:REV_VERB}
\end{table}

\section{Results}
\label{sec:results}
 
We fit all models to the $3-79 $ keV FPMA and FPMB spectra simultaneously using \textsc{xspec}. We first fit the \textsc{relxill} model described by Equation (\ref{eqn:prevxspecmodel}), and obtain results consistent with Z19 (see Table \ref{tab:previous}). We see that the best fitting spin parameter is very high, in agreement with Z19 and M13. The emissivity indices are also consistent with the Z19 analysis: $q_{\rm in}$ and $r_{\rm br}$ are small and $q_{\rm out}$ is large. This means that the Z19 model includes next to no contribution from intermediate disc radii, which cannot be reproduced in a general relativistic ray tracing model. The M13 analysis instead yielded a very high $q_{\rm in}$, and $q_{\rm out}\approx 0$. The high inner index can be reproduced in general relativity for a very compact source located very close to the black hole (\citealt{Wilkins2012,Dauser2013,Ingram2019}). However, $q_{\rm out} \approx 0$ is rather unphysical, requiring the irradiation to be $\sim$constant with radius despite the supply of gravitational potential energy dropping off as $\sim r^{-3}$.

We then fit the \textsc{reltrans} model (Equation \ref{eqn:xspecmodel}). We initially consider 6 different versions of this model. Table \ref{Tab: models} summarizes our naming convention: the labels A, B and C indicate the radial ionization profile used and the word in the subscript indicates the assumed abundances. We fix the spin to $a=0.998$ and leave $r_{\rm in}$ free. This choice maximises the range of $r_{\rm in}$ values that can be explored without $r_{\rm in}$ becoming smaller than the ISCO. The best fitting parameters and $\chi^2$ values for these 6 fits are presented in Table \ref{tab:results2}, and Fig. \ref{fig: resultsPlot} shows $\chi^2$ as a function of $r_{\rm in}$ for for these fits. We see that the constant ionization parameter models ($A_{\rm angr}$ and $A_{\rm wilm}$) return a higher value of $r_{\rm in}$ than the \textsc{relxill} model ($r_{\rm in}\approx 6.7~r_g$ compared with $r_{\rm in}=r_{\rm isco}$ for $a>0.989$; see Table \ref{tab:previous}). Introducing a more realistic ionization profile reduces the best fitting $r_{\rm in}$ and $\chi^2$, with the $C$ models returning the best fits. Comparison of the left and right hand panels of Fig. \ref{fig: resultsPlot} and perusal of Table \ref{tab:results2} confirm that the assumed abundances make little difference on the best fitting parameters (except for $N_H$). We therefore favour model $C_{\rm wilm}$, since the \cite{Wilms2000} abundances are the most up to date. This model still has a higher $\chi^2$ than the \textsc{relxill} model, but the emissivity profile is more physically plausible.

We then explore our favoured model further by additionally freeing $a$ in the fit. The model is sensitive to both $a$ and $r_{\rm in}$ because the spin influences the radial emissivity profile in the lamppost model. We do not consider spin constraints based on the emissivity profile to be robust because they are very dependent on the assumed coronal geometry, but freeing $a$ gives the model more freedom to explore parameter space. The parameters of this model
(model $C_{\rm wilm}^{'}$)
are quoted in the final row of Table \ref{tab:results2} and the observed spectrum unfolded around this model is shown in Fig. \ref{fig: bestFitSpectrum}. The improvement in $\chi^2$ over model $C_{\rm wilm}$ is not significant, but freeing the spin parameter may add to parameter uncertainties. A full exploration of the parameter space, obtained using a Markov Chain Monte Carlo (MCMC) simulation, is presented in Fig. \ref{fig:corner}. The only strong correlation between parameters is the anti-correlation between $r_{\rm in}$ and spin. This is simply because we prevented $r_{\rm in}$ from ever being smaller than the ISCO in the fit by expressing it in units of ISCOs. We then converted back to units of $r_g$ for Table \ref{tab:results2} and Fig. \ref{fig:corner}. $\log\xi_{\rm max}$ also weakly correlates with $a$ and anti-correlates with $r_{\rm in}$. These correlations are perhaps unsurprising, since we have already seen that the assumed ionization profile influences the best fitting $r_{\rm in}$ value.

A comparison of models $A_{\rm wilm}$, $B_{\rm wilm}$ and $C_{\rm wilm}^{'}$ is presented in Fig. \ref{fig:eemo_compare}. We see that the differences between the three models are very subtle. Yet the difference in the fit quality is large: model $C_{\rm wilm}$ is preferred to model $B_{\rm wilm}$ with $>3.38 \sigma$ confidence and to model $A_{\rm wilm}$ with $>4.3 \sigma$ confidence (lower limits are calculated from $\Delta\chi^2$ for one degree of freedom, but the models have the same number of degrees of freedom). We see from Fig. \ref{fig:eemo_compare} that the models with an ionization gradient have a greater contribution from the distant reflector than the constant ionization model. Table \ref{tab:REV_VERB} lists the \textsc{reltrans} reflection fraction for the best fitting parameters for each model. We quote both the \textit{system reflection fraction}, which is the fraction of coronal photons that hit the disc (this is exactly what \citealt{Dauser2016} refer to as the reflection fraction), and the \textit{observer's reflection fraction}, which is the ratio of bolometric reflected to bolometric direct flux that would be observed if the disc reflected like a perfect mirror (see \citealt{Ingram2019} for an extended discussion of these two definitions). We see that the models with a radial ionization profile have more reflection in the spectrum. The line feature is broader in these models, and so a similar spectral shape is achieved to the single ionization model for a higher reflection fraction. Note that the observer's reflection fraction depends on inclination angle but the system reflection fraction does not. Therefore some models have similar system reflection fractions but different observer's reflection fractions (e.g. $A_{\rm angr}$ and $A_{\rm wilm}$) because they differ in inclination angle.

In our best fitting model, the spin is high ($a=0.976$). However, the best fitting value of $r_{\rm in}$ provides a more robust lower limit for the spin, since this must presumably be $\geq r_{\rm isco}$. Taking the $1\sigma$ upper limit of $r_{\rm in} \leq 2.06$ therefore gives a $1\sigma$ lower limit on the spin of $a \geq 0.935$. We note however that the best fitting value of $r_{\rm in}$ can depend sensitively on the assumed model. As an example, we quote the results of fitting a simpler $\textsc{constant} \times \textsc{tbabs} \times \textsc{reltrans}$ model in Table \ref{tab:Z}. We see that this model, which does not include an intrinsic disc or distant reflector component and assumes ionization profile C, has a very large best fitting disc inner radius, $r_{\rm in} \approx 22~r_g$. This model provides a much worse fit to the data than our best fitting model and the inclination angle is lower than expected, but it is striking that our conclusions may have been so different had we not included the two extra spectral components in our model.

\section{Discussion}
\label{sec:discussion}

We have fit the spectrum of a 2012 \textit{NuSTAR} observation of GRS 1915+105 with a relativistic reflection model, employing three different assumptions for the radial dependence of the disc ionization parameter. We find that the most physically plausible ionization profile, with the disc density based on the \cite{Shakura1973} model, returns the best fit. We also find that the inferred disc inner radius is influenced by the assumed ionization profile.

Our best fitting model ($C_{\rm wilm}^{'}$; see Table \ref{tab:results2}) assumes the disc density corresponding to `zone A' of the \cite{Shakura1973} disc model, in which pressure and opacity are respectively dominated by radiation and electron scattering. We can check if this is a reasonable assumption by calculating the zone A outer radius $r_{\rm ab}$ (Equation 2.17 of \citealt{Shakura1973}, but note that they express this distance in units of $6 GM/c^2$). This depends on the viscosity parameter $\alpha$, the mass accretion rate in units of the Eddington limit $\dot{m}\equiv \dot{M}/\dot{M}_{\rm edd}$ and the black hole mass $M$. We assume $M=12.4~M_\odot$ \citep{Reid2014} and we can estimate $\dot{m}$ from the best fitting \textsc{reltrans} model (accounting for all relativistic effects) and the distance to the source, $D$. For $D=8.6$ kpc \citep{Reid2014}, the lamppost source luminosity of our best fitting model is $L_{\rm s} \approx 0.2~L_{\rm edd}$. Since there is presumably a second, equally powerful lamppost source that we cannot see on the far side of the disc, we estimate $\dot{m}\approx 0.4$. This implies that $r_{\rm ab} \approx 330-525~r_g$ for $\alpha$ in the range $0.01-1$, indicating that we do indeed expect the region of the disc that dominates the reflection emissivity to be comfortably in the zone A regime. Outside of the radius $r_{\rm ab}$ lies zone B, in which gas pressure dominates according to the \cite{Shakura1973} model.

\begin{figure}
    \includegraphics[width=\columnwidth,trim=1.8cm 1.6cm 2.0cm 11.0cm,clip=true]{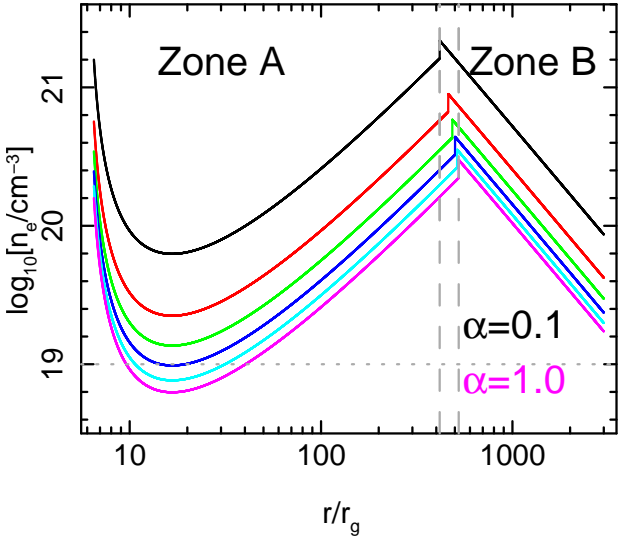}
    \vspace{-5mm}
    \caption{Electron density as a function of radius predicted by the standard disc model for values of $\alpha$ ranging in linear steps from $\alpha=0.1$ (black, top line) to $\alpha=1$ (magenta, bottom line). We assume $M=12.4~M_\odot$ and $\dot{m}=0.4$. The small discontinuity at the zone A to B transition occurs because pressure is assumed to be radiation dominated for $r<r_{\rm ab}$ and gas dominated for $r>r_{\rm ab}$, whereas in reality both contributions are important for $r\sim r_{\rm ab}$. The dotted line depicts the density inferred from our best fitting model.}
\label{fig:density}
\end{figure}

The peak disc ionization of our best fitting model is rather high ($\log\xi_{\rm max}=4.18_{-0.3}^{+0.3}$). We can calculate the disc density implied by our fit by rearranging Equation (11) from \cite{Mastroserio2019}  to find $n_e \approx 10^{19} {\rm cm}^{-3}$ at the radius of peak ionization (assuming the same mass and distance as before). We can check if this value is reasonable: Fig. \ref{fig:density} shows $n_e$ as a function of radius for the \cite{Shakura1973} disc model with 6 different values of $\alpha$ in the range $0.1-1$ in linearly spaced steps (calculated using their equations 2.11 and 2.16). The dashed lines mark the maximum and minimum $r_{\rm ab}$ transition radius for the $\alpha$ values considered. We see that densities as low as $n_e \sim 10^{19} {\rm cm}^{-3}$ are possible in the standard disc model for $\alpha \gtrsim 0.6$, although this is at $r\gtrsim 10~r_g$, whereas the peak ionization in our fit is at $r\sim~2.6~r_g$. This is a higher value of $\alpha$ than is
typically assumed,
but departures from the simple standard disc model (e.g. significant magnetic pressure) could allow lower densities. Moreover, such low values for $n_e$ have recently been inferred observationally \citep{Tomsick2018,Jiang2019}.

The high ionization parameter associated with the \textsc{xillver} component ($\log\xi_{\rm xill}=3.16_{-0.14}^{+0.08}$) also requires a low disc density. Taking the extreme assumption that the surface of the disc in the region that produces the \textsc{xillver} component exactly faces the source -- corresponding to a heavily flared or warped accretion disc -- gives an upper limit on the density of $\log(n_e/{\rm cm}^{-3}) \lesssim 19 - 2\log[r_{\rm xill}/(100~r_g)]$, where $r_{\rm xill}$ is the characteristic radius from where the \textsc{xillver} component is emitted. \footnote{Note, $\log$ is taken to mean logarithm to the base 10 throughout.} Therefore for the characteristic radius to be large enough for the rotational broadening of the line to be small ($r_{\rm xill} \gtrsim 100~r_g$), the density must be $n_e \lesssim 10^{19} {\rm cm}^{-3}$ and the disc must be flared or warped. This too is a lower density than is predicted by the \cite{Shakura1973} model. A more sophisticated treatment would be to include a more realistic disc scale height in the relativistic reflection model (e.g. \citealt{Taylor2018}), which should dispense with the need for the extra \textsc{xillver} component. Alternatively, the additional \textsc{xillver} component could include a contribution from another structure such as a wind or the companion star surface, although the latter is unlikely for GRS 1915+105 owing to its very large binary separation. We also note that the \textsc{xillver} calculation we use for both the relativistic and distant reflectors is for a fixed disc density of $n_e=10^{15} {\rm cm}^{-3}$. Using a reflection model that explicitly includes $n_e$ as a parameter in future will therefore improve the physical consistency of our fits.

\section{Conclusions}
\label{sec:conclusions}

We have modelled the X-ray reflection spectrum of GRS 1915+105 by including a self-consistently calculated radial profile of the disc ionization parameter. Our best fit is achieved by assuming the disc to be in the radiation pressure dominated zone A regime of the \cite{Shakura1973} disc model, similar to the results of \cite{Mastroserio2019} for Cygnus X-1. Since the accretion rate during this observation was high, we expect at least the inner $\sim 300~r_g$ of the disc to be in the zone A regime. It is encouraging that it seems possible to infer the radial disc structure from the reflection spectrum. In future it may be possible to use the best fitting ionization profile to infer the point during an outburst that the inner disc transitions from being gas to radiation pressure dominated.

The inferred disc inner radius, and by extension the inferred black hole spin, depends on the assumed ionization profile. In our case, the constant ionization model has a disc inner radius of $r_{\rm in} \approx 6.7~r_g$ whereas the zone A model has $r_{\rm in} \approx 1.7~r_g$. This corresponds to a large change in the inferred spin if $r_{\rm in}$ is assumed to be at the ISCO. It is therefore possible that the inclusion of a self-consistent radial ionization profile will change the inferred spin values of AGN and XRBs. Although the inferred spin is effectively increased in this case, it is not clear if the inferred spin will always systematically increase. Use of a radial ionization profile would likely also impact on the parameterized deformations to the Kerr metric explored by Z19.

This observation exhibits a Type-C QPO at a frequency of $\nu_{\rm qpo}\sim 1.5$ Hz \citep{Bachetti2015}. It is difficult to reconcile this QPO frequency with our small best fitting disc inner radius for any existing QPO model (Ingram \& Motta submitted), since they all interpret the observed increase in $\nu_{\rm qpo}$ as a decrease in $r_{\rm in}$. Precession of a vertically extended corona at the jet base \citep{Liska2018} may be able to reproduce the observed QPO frequencies for small disc inner radii whilst remaining consistent with the observed QPO phase dependence of the iron line profile in H 1743-322 \citep{Ingram2016,Ingram2017}. Alternatively, future inclusion of more sophisticated assumptions in the reflection and/or continuum model may push out the inferred $r_{\rm in}$ enough to relieve some of the tension with existing QPO models.


\section*{Acknowledgements}

SS and AI acknowledge support from the Royal Society. We acknowledge useful comments from the referee Javier Garcia that improved the paper.


\bibliographystyle{mnras}
\bibliography{biblio} 




\bsp	
\label{lastpage}
\end{document}